\documentclass[12pt,a4paper,dvips,twoside]{article}
\title{} 
\author{} 
\date{} 

\usepackage{amsfonts}
\usepackage{amssymb}
\usepackage{times}
\usepackage{graphicx}

\def\onehalf{{\scriptstyle\frac{1}{2}}}

\let\ro=\varrho

\begin{document}

\title{An iterative algorithm for evaluating approximations
to the optimal exercise boundary for a nonlinear Black--Scholes equation}

\author{Daniel \v Sev\v covi\v c \thanks{
Department of Applied Mathematics and Statistics,
Faculty of Mathematics, Physics \& Informatics, Comenius University, 
842 48 Bratislava, Slovak Republic, \quad {\tt sevcovic@fmph.uniba.sk}
\vskip 0mm
This work was supported by VEGA grant 1/3767/06.}
}


\maketitle

\begin{abstract}
The purpose of this paper is to analyze and compute the early exercise boundary  
for a class of nonlinear Black--Scholes equations with a nonlinear volatility which 
can be a function of the second derivative of the option price itself. A motivation 
for studying the nonlinear Black--Scholes equation with a nonlinear volatility 
arises from option pricing models taking into account e.g. nontrivial transaction costs, 
investor's preferences, feedback and illiquid markets effects and risk from a volatile (unprotected) portfolio. 
We present a new method how to transform the free boundary problem for the early exercise boundary 
position into a solution of a time depending nonlinear parabolic equation defined on a fixed domain. 
We furthermore propose an iterative numerical scheme that can be used to find 
an approximation of the free boundary. We present results of numerical
approximation of the early exercise boundary  
for various types of nonlinear Black--Scholes equations and we discuss
dependence of the free boundary on various model parameters. 
\end{abstract}



\section{Introduction}

In the past years, the Black--Scholes equation for pricing derivatives has attracted a lot of 
attention from both theoretical as well as practical point of view. Recall that a European Call 
(Put) option is the right but not obligation to purchase (sell) an underlying  asset at the expiration 
price $E$ at the expiration time $T$. In an idealized  financial market the 
price of an option can be computed from the well-known Black--Scholes equation 
derived by Black and Scholes in \cite{BS}, and, independently by Merton 
(see also \cite{Kw}, Dewynne {\em et al.} \cite{DH}, Hull \cite{H}).
Assuming that the underlying asset follows a geometric Brownian motion one can derive a governing 
partial differential equation for the price of an option. We remind ourselves that the equation 
for option's price $V(S,t)$ is the following parabolic PDE:
\begin{equation}
\partial_t V +  (r-q) S\partial_S V + \onehalf \sigma^2 S^2 \partial_S^2 V - r V =0
\label{BS}
\end{equation}
where $\sigma$ is the volatility of the underlying asset price process, $r>0$ is the interest 
rate of a zero-coupon bond, $q\ge 0$ 
is the dividend yield rate. A solution $V=V(S,t)$ represents the price of an option at 
time $t\in[0,T]$ if the price of an underlying asset is $S>0$. 
In this paper we shall focus our attention to the case when the diffusion coefficient
$\sigma^2$ may depend on the time $T-t$ to expiry, the asset price $S$ 
and the second derivative $\partial^2_S V$ of the option price  (referred to as $\Gamma$), i.e.
\begin{equation}
\sigma = \sigma(S^2\partial^2_S V, S, T-t)\,.
\label{sigma}
\end{equation}
A motivation for studying the nonlinear Black--Scholes equation (\ref{BS})
with a volatility $\sigma$ having a general form (\ref{sigma}) arises from option pricing 
models taking into account nontrivial transaction costs, market feedbacks and/or risk from a volatile 
(unprotected) portfolio. Recall that the linear Black--Scholes equation with $\sigma$ constant has 
been derived under several restrictive assumptions like e.g. frictionless, liquid, complete markets, etc. 
We also recall that the linear Black--Scholes equation provides a perfectly replicated hedging portfolio. 
In recent years, some of these assumptions have been relaxed in order to model, for instance, the 
presence of transaction costs (see e.g. Leland \cite{Le}, Hoggard {\em et al.} \cite{HWW}, Avellaneda and 
Paras \cite{AP}), feedback and illiquid market effects due to large traders choosing given stock-trading 
strategies (Frey and Patie \cite{FP}, Frey and Stremme \cite{FS}, During {\em et al.}\cite{DFJ},
Sch\"onbucher and  Wilmott \cite{SW}), imperfect replication and investor's preferences (Barles and Soner 
\cite{BaSo}), risk from unprotected portfolio (Kratka \cite{Kr}, Janda\v{c}ka and \v{S}ev\v{c}ovi\v{c}
\cite{JS}). One of the first nonlinear models is the so-called Leland model (c.f. \cite{Le}) for pricing Call 
and Put options under the presence of transaction costs. It has been generalized for more complex 
options  
by Hoggard, Whaley and Wilmott  in \cite{HWW}. In this model the volatility $\sigma$ is given by
$\sigma^2 (S^2\partial^2_S V, S, \tau) = \hat\sigma^2 ( 1 + \hbox{Le}\, \hbox{sgn}(\partial^2_S V) )$ where
$\hat\sigma>0$ is a constant historical volatility of the underlying asset price process
and $\hbox{Le}>0$ is the so-called Leland constant given by 
$\hbox{Le} = \sqrt{2/\pi} C/(\hat\sigma \sqrt{\Delta t})$
where $C>0$ is a constant round trip transaction cost per unit dollar of transaction in the assets market
and $\Delta t>0$ is the time-lag between portfolio adjustments. 

If transaction costs are taken into account perfect replication of the
contingent claim is no longer possible and further restrictions are needed 
in the model. By assuming that investor's preferences are characterized by an 
exponential utility function   Barles and Soner (c.f. \cite{BaSo}) derived a 
nonlinear Black--Scholes equation with the volatility $\sigma$ given by
\begin{equation}
\sigma^2(S^2\partial^2_S V, S, \tau) 
= \hat\sigma^2 \left(1+\Psi(a^2 e^{r \tau} S^2\partial^2_S V)\right)
\label{barles}
\end{equation}
where $\Psi$ is a solution to the 
ODE: $\Psi^\prime(x) = (\Psi(x)+1)/(2 \sqrt{x\Psi(x)} -x), \Psi(0)=0,$
and $a>0$ is a given constant representing risk aversion. 
Notice that $\Psi(x)=O(x^\frac{1}{3})$ for 
$x\to 0$ and $\Psi(x)=O(x)$ for $x\to\infty$. 

Another popular model has been derived for the case when the asset dynamics 
takes into account the presence of feedback and illiquid market effects. Frey and Stremme 
(c.f. \cite{FS,FP}) introduced directly the asset price dynamics in the case
when a large trader chooses a given stock-trading strategy (see also \cite{SW}). 
The diffusion coefficient $\sigma$ is again nonconstant and it can be expressed as:
\begin{equation}
\sigma^2(S^2\partial^2_S V, S, \tau) = \hat\sigma^2 \left(1-\varrho \lambda(S) S\partial^2_S V\right)^{-2}
\label{frey}
\end{equation}
where $\hat\sigma^2, \varrho>0$ are constants and $\lambda(S)$ is a strictly convex function, 
$\lambda(S)\ge 1$.

The last example of the Black--Scholes  equation with a nonlinearly depending volatility
is the so-called Risk Adjusted Pricing Methodology model proposed by Kratka in \cite{Kr}
and revisited by Janda\v{c}ka and \v{S}ev\v{c}ovi\v{c} in \cite{JS}. 
In order to maintain (imperfect) replication of a portfolio by the delta hedge one has to make 
frequent portfolio adjustments leading to a substantial increase in transaction costs. 
On the other hand, rare portfolio adjustments may lead to an increase of the risk arising 
from a volatile (unprotected) portfolio. In the RAPM model the aim is to 
optimize  the time-lag $\Delta t$ between consecutive portfolio adjustments. 
By choosing $\Delta t>0$ in such way that the sum of the rate of transaction costs
and the rate of a risk from unprotected portfolio is minimal 
one can find the optimal time lag $\Delta t>0$ (see \cite{JS} for details). In the
RAPM model, it turns out that the volatility is again nonconstant and 
it has the following form:
\begin{equation}
\sigma^2(S^2\partial^2_S V, S, \tau) = \hat\sigma^2 \left(1 + \mu (S\partial^2_S V)^{\frac{1}{3}} \right)\,. 
\label{RAPM}
\end{equation}
Here $\hat\sigma^2>0$ is a constant and $\mu=3(C^2 R/2\pi)^{\frac{1}{3}}$ where $C, R\ge 0$ are 
nonnegative constant representing the transaction cost measure and the risk premium measure, resp. 
(see \cite{JS} for details).

Notice that all the above mentioned nonlinear models are consistent with the original 
Black--Scholes equation in the case the additional model parameters (e.g. Le, $a$, 
$\varrho$, $\mu$)
are vanishing. If plain Call or Put vanilla options are concerned then the function $V(S,t)$ is convex
in $S$ variable and therefore each of  the above mentioned models 
has a diffusion coefficient strictly larger than $\hat\sigma^2$ leading to a larger values of computed option 
prices. They can be therefore identified with higher Ask option prices, i.e. offers to sell an option. 
Furthermore, these models have been considered and analyzed mostly for European options, i.e. 
options can be exercised only at the maturity $t=T$. On the other hand, American  options 
are more common in  financial markets as they allow
for exercising of an option anytime before the expiry $T$. In the case of an American Call option 
a solution to equation (\ref{BS}) is defined on a time dependent domain $0<t<T,\ 0<S<S_f(t)$. 
It is subject to the boundary conditions
\begin{equation}
V(0,t) = 0\,,\ \ V(S_f(t), t)= S_f(t)-E\,, \ \ \partial_S V (S_f(t),t) = 1\,,
\label{bcCall}
\end{equation}
and terminal pay-off condition at expiry $t=T$
\begin{equation}
V(S,T)=\max(S-E, 0)
\label{tcCall}
\end{equation}
where $E>0$ is a strike price (c.f. \cite{DH,Kw}). 
One of important problems in this field is the analysis of the early exercise boundary 
$S_f(t)$ and  the optimal stopping time (an inverse function to $S_f(t)$) 
for American Call options on stocks paying a continuous dividend $q>0$. 
However, an exact analytical expression for the free boundary profile is not even known
for the case when the volatility $\sigma$ is constant. Many authors have investigated 
various approximation models leading to approximate expressions for valuing American Call and
Put options: analytic approximations 
(Barone--Adesi and Whaley \cite{BW}, 
Kuske and Keller \cite{KK},
Dewynne {\em et al.} \cite{DH},
Geske {\em et al.} \cite{GJ,GR},
MacMillan \cite{Mac},
Mynemi \cite{M}); 
methods of reduction to a nonlinear integral equation 
(Alobaidi \cite{A}, 
Kwok \cite{Kw},
Mallier {\em et al.} \cite{Ma,MA},
\v{S}ev\v{c}ovi\v{c} \cite{Se},
Stamicar {\em et al.} \cite{SSC}). Concerning numerical methods for solving free boundary problem we refer to
the book by Kwok \cite{Kw} and recent papers by Ehrhardt and Mickens \cite{EM} and Zhao {\em et al.} \cite{ZCD}.

The main goal of this paper is to propose a new iterative numerical algorithm for solving the free boundary
problem for an American Call option in the case the volatility $\sigma$ may depend 
on the option and asset values as well as on the time $T-t$ to expiry. The key idea of the
proposed algorithm consists in transformation of the free boundary problem into a semilinear 
parabolic equation 
defined  on a fixed spatial domain coupled with a nonlocal algebraic constraint 
equation for the free boundary position. 
Since the resulting parabolic equation contains a strong convective term we make use of the operator
splitting method in order to overcome numerical difficulties. 
Full space-time discretization of the problem leads to a system 
of semi-linear algebraic equations that can be solved by an iterative procedure at each time level. 

The paper has the following plan:
in the next section we transform the free boundary problem into a system consisting of 
a nonlinear parabolic equation defined on a fixed domain with time depending coefficients
and an algebraic constraint equation for the free boundary position. In section 3 we present a numerical 
discretization scheme based on the idea of operator splitting. In the last section 4 we present 
several numerical results for nonlinear Black--Scholes equations with volatility functions $\sigma$ defined 
in (\ref{barles}) and \ref{RAPM}). We make a comparison to well-known methods in the case the volatility 
$\sigma$ is constant. Finally, we discuss dependence of the free boundary position with respect to various
parameters entering expressions  (\ref{barles}) and (\ref{RAPM}).

\section{Landau fixed domain transformation}

The main goal of this section is to perform a fixed domain transformation (referred to as 
Landau's transformation) of the free boundary problem for the nonlinear Black--Scholes 
equation (\ref{BS}) into a parabolic equation defined on a fixed spatial domain. 
For the sake of simplicity we will present a detailed derivation of an equation only 
for the case of an American Call option. Derivation of the corresponding equation for 
the American Put option is similar.

Let us consider the following change of variables:
\[
\tau= T-t, \quad x=\ln\left(\ro(\tau)/S\right)
\ \ \hbox{where}\ \ \ro(\tau)=S_f(T-\tau).
\]
Then $\tau\in(0,T)$ and $x\in(0,\infty)$ iff $S\in(0,S_f(t))$. The value $x=0$ corresponds to
the free boundary position $S=S_f(t)$ whereas $x\approx +\infty$ corresponds to the default
value $S=0$ of the underlying asset. Following Stamicar {\em et al.}  \cite{SSC} and 
\v{S}ev\v{c}ovi\v{c} \cite{Se}
we construct the so-called synthetic portfolio function $\Pi=\Pi(x,\tau)$ 
defined as follows:
\begin{equation}
\Pi(x,\tau)= V(S,t) - S \partial_S V(S,t)\,.
\label{transformacia}
\end{equation}
It corresponds to a synthetic portfolio consisting of one long positioned option and
$\Delta= \partial_S V$ underlying short stocks. Clearly, we have
\[
\partial_x\Pi = S^2 \partial_S^2 V,  \ \ \partial_\tau \Pi + \frac{\dot\ro}{\ro}\partial_x\Pi =
- \partial_t \left( V - S \partial_S V\right)
\]
where we have denoted $\dot\ro=d\ro/d\tau$. 
Assuming sufficient smoothness of a solution $V=V(S,t)$ to (\ref{BS}) 
we can deduce from (\ref{BS}) a parabolic equation for the synthetic portfolio 
function $\Pi=\Pi(x,\tau)$
\[
\partial_\tau\Pi  + (b(\tau) - \onehalf \sigma^2 ) \partial_x \Pi 
- \onehalf \partial_x\left( \sigma^2 \partial_x\Pi \right) + r \Pi=0
\]
defined on a fixed domain $x\in R,\, t\in(0,T),$ with a time-dependent coefficient
\begin{equation}
b(\tau)= {\dot\ro(\tau)\over\ro(\tau)} + r -q 
\label{bfun}
\end{equation}
and a diffusion coefficient given by: 
$\sigma^2=\sigma^2( \partial_x\Pi(x,\tau), \ro(\tau)e^{-x}, \tau)$
depending on $\tau, x$ and the gradient $\partial_x\Pi$ 
of a solution $\Pi$. Now the boundary conditions $V(0, t)=0, V(S_f(t), t)= S_f(t) - E$ 
and $\partial_S V(S_f(t), t)= 1$ imply
\begin{equation}
\Pi(0, \tau) = -E, \quad \Pi(+\infty, \tau) =0\,, \quad 0<\tau<T\,,
\end{equation}
and, from the terminal pay-off diagram for $V(S,T)$, we deduce
\begin{equation}
\Pi(x,0)=
\left\{
   \begin{array}{lll}
-E & \hbox{for} \ x<\ln\left({\ro(0)\over E}\right)\hfil
\\
\ \ 0 & \hbox{otherwise.}\hfil
\end{array}
\right.
\end{equation}
In order to close up the system of equations that determines the value of 
a synthetic portfolio $\Pi$ we have to construct an equation for the free
boundary position $\ro(\tau)$. Indeed, both the coefficient $b$ as well as the
initial condition $\Pi(x,0)$ depend on the function $\ro(\tau)$. 
Similarly as in the case of a constant volatility $\sigma$ (see \cite{Se,SSC})
we proceed as follows: since $S_f(t)-E=V(S_f(t),t)$ and $\partial_S V(S_f(t),t)=1$ we have
${\scriptstyle \frac{d}{dt}} S_f(t) 
= \partial_S V(S_f(t),t) {\scriptstyle \frac{d}{dt}} S_f(t) + \partial_t V(S_f(t), t)$
and so $\partial_t V(S,t)=0$ along the free boundary $S=S_f(t)$. Moreover, assuming $\partial_x\Pi$ is 
continuous  up to the boundary $x=0$ we obtain $S^2\partial^2_S V(S,t) \to  \partial_x\Pi(0,\tau)$
and $S\partial_S V(S,t)\to \ro(\tau)$ as $S\to S_f(t)^-$. Now, by taking the limit 
$S\to S_f(t)^-$ in the Black--Scholes equation (\ref{BS}) we obtain
$(r-q)\ro(\tau) +\onehalf \sigma^2 \partial \Pi(0,\tau) -r(\ro(\tau)-E) =0$.
Therefore
\[
\ro(\tau)= {rE\over q} 
+ \frac{1}{2q} \sigma^2(\partial_x\Pi(0,\tau), \ro(\tau), \tau)
\partial_x\Pi(0,\tau)
\]
for $0<\tau\le T$.
Recall that in the linear case when $\sigma>0$ is constant the initial position 
of the interface $\ro(0)$ is given by $\ro(0)=rE/q$ if $r\ge q>0$ and $\ro(0)=E$
if $0<r<q$ (see Dewynne {\em et al}.~\cite{DH} or \v{S}ev\v{c}ovi\v{c} \cite{Se}). 
We also recall
that the value of $\ro(0)$ in the case $r\ge q>0$ can derived easily from the smoothness
assumption made on $\partial_x\Pi$ at the origin $x=0, \tau=0$. Indeed, continuity of 
$\partial_x\Pi$ at the origin $(0,0)$ implies $\lim_{\tau\to0^+}\partial_x\Pi(0,\tau)
=\partial_x\Pi(0,0) =\lim_{x\to0^+}\partial_x \Pi(x,0) = 0$
because  $\Pi(x,0)=-E$ for $x$ close to $0^+$. From the above equation 
for $\ro(\tau)$ we deduce $\ro(0)= {rE\over q}$  by taking the limit $\tau\to 0^+$. Henceforth, 
we restrict our attention to the case when the interest rate is greater 
then the dividend yield rate, i.e.
\begin{equation}
0<q\le r
\label{rD}
\end{equation}
leading to the initial position of the free boundary $\ro(0)=rE/q$.
Putting all the above equations together we end up with a closed system
of equations for $\Pi=\Pi(x,\tau)$ and $\ro=\ro(\tau)$
\begin{eqnarray}
&&\partial_\tau\Pi  + (b(\tau) -\onehalf \sigma^2 )\partial_x \Pi
- \onehalf \partial_x\left( \sigma^2 \partial_x\Pi\right) + r \Pi=0\,,
\nonumber \\
&&\Pi(0, \tau) = -E\,, \ \  \Pi(+\infty, \tau) =0\,,\  x>0\,, \tau\in(0,T)\,,
\nonumber\\
&&\Pi(x,0)= \left\{
\begin{array}{lll}
-E & \quad\hbox{for} \ x<\ln(r/q)\hfil
\\
\ \ \  0 & \quad\hbox{otherwise\,,}\hfil
\end{array}
\right.
\label{tBS}
\end{eqnarray}
where
$\sigma=\sigma(\partial_x\Pi(x,\tau), \ro(\tau)e^{-x},\tau )\,,\ 
b(\tau)= {\dot\ro(\tau)\over\ro(\tau)} + r - q$ and the free boundary position 
$\ro(\tau)=S_f(T-\tau)$ satisfies an implicit algebraic equation
\begin{equation}
\ro(\tau)={r E\over q} + {\sigma^2(\partial_x\Pi(0,\tau), \ro(\tau), \tau )\over 2q}
\partial_x\Pi(0,\tau)\,,\quad \hbox{with }\ \ro(0)={rE\over q}
\label{ro}
\end{equation}
where $\tau\in(0,T)$.
Notice that, in order to guarantee parabolicity of equation (\ref{tBS}) we have
to assume that the function $p\mapsto \sigma^2(p, \ro(\tau)e^{-x}, \tau) p$ is strictly
increasing. More precisely, we shall assume that there exists a positive constant $\gamma>0$
such that 
\begin{equation}
\sigma^2(p, \xi, \tau)  + p\partial_p\sigma^2(p, \xi, \tau)\ge \gamma>0
\label{strict} 
\end{equation}
for any $\xi>0, \tau\in(0,T)$ and $p\in R$.

\medskip
\noindent{\bf Remark 1.}
In \cite{Se} the author derived a single equation for the position of the free boundary 
$\ro$ for the case when the volatility $\sigma=\hat\sigma$ is constant. In this case one can solve 
the initial-boundary
value problem for the linear parabolic equation (\ref{tBS}) with spatially independent coefficients
by means of one-sided sine and cosine Fourier transforms in the spatial $x$ variable. 
The explicit formula for $\Pi(x,\tau)$
together with equation (\ref{ro}) enables us to conclude that the free boundary position $\ro$
satisfies the following nonlinear weakly singular integral equation
\begin{eqnarray}
\ro(\tau) &=& \frac{r E}{q} \biggl( 1+ \frac{\hat\sigma}{r\sqrt{2\pi\tau}}
\ \exp\left(-r\tau - (A_{\tau,0}+\ln(r/q))^2/(2\hat\sigma^2\tau)\right)
\nonumber\\
&& + \frac{1}{\sqrt{2\pi}}\int_0^\tau
\biggl[\hat\sigma+\frac{1}{\hat\sigma}\biggl(1-\frac{q\ro(s)}{r E} \biggr)
{A_{\tau,s}\over\tau-s}
\biggr]
\frac{\exp\left(-r(\tau-s) -\frac{A_{\tau,s}^2}{2\hat\sigma^2(\tau-s)}\right)}{\sqrt{\tau-s}}
\, \hbox{d}s \biggr)
\label{integral}
\end{eqnarray}
where the function $A_{\tau,s}$ depends upon $\ro$ via $A_{\tau,s} = \ln\ro(\tau) -\ln\ro(s)
+\left(r-q-\onehalf\hat\sigma^2\right)(\tau-s)$.
The above integral equation can be solved by an iterative procedure based on a
Banach fixed point argument (see \cite{Se} for details). 
It is worthwhile noting that this approach cannot be applied to the case when the 
volatility may depend on the asset price $S$ and/or the second derivative of the option price
as the Fourier integral transform technique is no longer applicable. However, in section 4 
we shall use a solution computed from
the nonlinear integral equation (\ref{integral}) in order to make comparison of our iterative 
numerical scheme in the case when the volatility $\sigma$ is constant.

\medskip
\noindent{\bf Remark 2.}
We also present a formula for pricing American
Call options based on the solution $(\Pi,\ro)$ to (\ref{tBS})--(\ref{ro}).
By (\ref{transformacia}) we have
$
\partial_S \left( S^{-1} V(S,t) \right) =
- S^{-2} \Pi\left( \ln\left(\ro(T-t)/S\right), \, T-t \right)$.
Taking into account the boundary condition $V(S_f(t), t) = S_f(t)-E$ and
integrating the above equation from $S$ to $S_f(t)=\ro(T-t)$, we obtain
the expression for the option price $V(S,t)$:
\begin{equation}
V(S,T-\tau)= {S\over\ro(\tau)} \left(
\ro(\tau) -E + \int_0^{\ln{\ro(\tau)\over S}} {\rm e}^x \Pi(x,\tau) \, dx
\right)\,.
\end{equation}

\section{Discretization scheme. An iterative algorithm for approximation of the early exercise 
boundary}

In this section we derive a full space-time discretization scheme for a numerical approximation
of the problem (\ref{tBS}), (\ref{ro}). Recall that in the case of a constant volatility
there are, in principle, two ways how to solve numerically the free boundary problem
for the value of an American Call resp. Put option and the position of the early exercise 
boundary. The first class of algorithms is based on reformulation of the problem
in terms of a variational inequality (see Kwok \cite{Kw} and references therein). The variational
inequality can be then solved numerically by the so-called Projected Super Over Relaxation method
(PSOR for short). An advantage of this method is that it gives us immediately the value of a
solution. A disadvantage is that one has to solve large systems of linear equations 
iteratively taking into account the obstacle for a solution, and, secondly, the free 
boundary position should be deduced from the solution a posteriori. Moreover, the PSOR method
is not directly applicable for solving the problem (\ref{BS})-(\ref{bcCall}) when the diffusion
coefficient $\sigma$ may depend on the second derivative of a solution itself. The second
class of methods is based on derivation of a nonlinear integral equation for
the position of the free boundary  without the need of knowing the option price itself
(see e.g. Kuske-Keller \cite{KK}, Mallier and Alobaidi \cite{A,Ma,MA}, 
\v{S}ev\v{c}ovi\v{c} {\em et al.}
\cite{Se,SSC}. In this approach an advantage is that only a single equation for the free boundary
has to be solved; a disadvantage is that the method is based on integral
transformation techniques and therefore the assumption $\sigma$ is constant  
is crucial. 

In our approach we make an attempt to take advantages of both above mentioned methods. As it
was mentioned in the previous section we are going to solve the system of nonlinear equations 
(\ref{tBS}) with constraint (\ref{ro}).
Because the volatility $\sigma$ may be nonconstant we cannot use integral transformation
techniques in order to derive a single integral equation for $\ro(\tau)$. However, the form
of the system (\ref{tBS}), (\ref{ro}) allows for an efficient and fast numerical
algorithm for computing of the early exercise boundary  position $\ro(\tau)=S_f(T-\tau)$.

The idea of the iterative numerical algorithm for solving the problem (\ref{tBS}), (\ref{ro})  
is rather simple: we use the backward Euler method of 
finite differences in order to discretize the parabolic equation (\ref{tBS}) in time. In each time
level we find a new approximation of a solution pair $(\Pi, \ro)$. First we determine a new position
of $\ro$ from the algebraic equation (\ref{ro}). 
We remind ourselves that (even in the case $\sigma$ is constant) the free boundary function $\ro(\tau)$ 
behaves like $r E/q + O(\tau^{1/2})$ for $\tau\to 0^+$ (see e.g. \cite{DH} or \cite{Se}) and so $b(\tau) = 
O(\tau^{-1/2})$. Hence the convective term $b(\tau)\partial_x\Pi$ becomes 
a dominant part of equation (\ref{tBS}) for small values of $\tau$. In order to overcome this difficulty we use the operator splitting
method for successive solving of the convective and diffusion parts of equation (\ref{tBS}). 
Since the diffusion coefficient depends on the solution $\Pi$ itself we make several micro-iterates
to find a solution of a system of nonlinear algebraic equations. 

Now we present our algorithm in more details. 
We restrict the spatial domain $x\in (0,\infty)$ to a finite interval of values
$x\in (0,L)$ where $L>0$ is sufficiently large. For practical purposes one can
take $L\approx 3$ as it corresponds to the interval  $S\in (S_f(t) e^{-L}, S_f(t))$
in the original asset price variable $S$. The value $S_f(t) e^{-L}$ is then a good 
approximation for the default value $S=0$ if $L\approx 3$. 
Let us denote by $k>0$ the time step, $k=T/m$, and, by $h>0$ the spatial step, $h=L/n$
where $m, n\in N$ stand for the number of time and space discretization steps, resp. 
We denote by $\Pi_i^j$ an approximation of $\Pi( x_i, \tau_j)$, $\ro^j \approx \ro(\tau_j)$,
$b^j \approx b(\tau_j)$ where $x_i=i h, \tau_j=j k$. We approximate the value of the 
volatility $\sigma$ at  the node $(x_i,\tau_j)$ by finite difference as follows:
\[
\sigma_i^j = \sigma_i^j(\ro^j, \Pi^j) =\sigma((\Pi_{i+1}^j - \Pi_i^j)/h, \ro^j e^{-x_i}, \tau_j)\,.
\]
Then for the backward in time Euler finite difference approximation of equation (\ref{tBS}) we have
\begin{equation}
\frac{\Pi^j-\Pi^{j-1}}{k} + \left( b^j -\frac12 (\sigma^j)^2 \right) \partial_x\Pi^j 
-\frac12 \partial_x \left( (\sigma^j)^2 \partial_x\Pi^j\right) +r \Pi^j = 0
\label{timediscret}
\end{equation}
and the solution $\Pi^j=\Pi^j(x)$ is subject to Dirichlet boundary conditions at $x=0$  and $x=L$. We set
$\Pi^0(x)=\Pi(x,0)$.
Now we decompose the above problem into two parts - a convection part and a diffusive part by introducing 
an auxiliary intermediate step $\Pi^{j-\onehalf}$:

({\it Convective part})
\begin{equation}
\frac{\Pi^{j-\onehalf}-\Pi^{j-1}}{k} + b^j  \partial_x\Pi^{j-\onehalf} = 0\,,
\label{convective}
\end{equation}

({\it Diffusive part})
\begin{equation}
\frac{\Pi^{j}-\Pi^{j-\onehalf}}{k}   - (\sigma^j)^2  \partial_x\Pi^j
-\frac12 \partial_x \left( (\sigma^j)^2 \partial_x\Pi^j\right) +r \Pi^j = 0\,.
\label{diffusion}
\end{equation}
The idea of the operator splitting technique now consists in comparison the sum of solutions to 
convective and diffusion 
part to a solution of (\ref{timediscret}). Indeed, if $\partial_x\Pi^j \approx \partial_x\Pi^{j-\onehalf}$ then 
it is reasonable to assume that $\Pi^j$ computed from the system (\ref{convective})--(\ref{diffusion})
is a good approximation of the system (\ref{timediscret}).

The convective part can be approximated by an explicit solution to the transport equation:
\begin{equation}
\partial_\tau\tilde\Pi + b(\tau) \partial_x\tilde\Pi =0 \qquad \hbox{for } x>0, \tau\in(\tau_{j-1},\tau_j] 
\label{convective-anal}
\end{equation}
subject to the boundary condition $\tilde\Pi(0,\tau) =-E$ and initial condition  
$\tilde\Pi(x,\tau_{j-1})=\Pi^{j-1}(x)$. Since the free boundary $\ro(\tau)=S_f(T-\tau)$ must be an increasing
function in $\tau$ and we have assumed $0<q\le r$ we have $b(\tau)=\dot\ro(\tau)/\ro(\tau) +r-q >0$ and
so the in-flowing boundary condition $\tilde\Pi(0,\tau) =-E$ is consistent with the transport equation.
Let us denote by $B(\tau)$ the primitive function to $b(\tau)$, i.e. $B(\tau)=\ln \ro(\tau) + (r-q)\tau$.
Equation (\ref{convective-anal}) can be integrated to obtain its explicit solution:
\begin{equation}
\tilde\Pi(x,\tau)=\left\{ 
\matrix{
\Pi^{j-1}(x-B(\tau)+B(\tau_{j-1})) \hfill & \quad \hbox{if } x-B(\tau)+B(\tau_{j-1})>0\,, \hfill\cr 
-E \hfill & \quad \hbox{otherwise.} \hfill
}
\right.
\label{convective-explicit}
\end{equation}
Thus the spatial approximation $\Pi^{j-\onehalf}_i$ can be constructed from the formula
\begin{equation}
\Pi^{j-\onehalf}_i=\left\{ 
\matrix{
\Pi^{j-1}(\xi_i) \hfill & \quad \hbox{if } \xi_i=x_i- \ln\ro_j + \ln\ro_{j-1} - (r-q)k>0\,, \hfill\cr 
-E \hfill & \quad \hbox{otherwise,} \hfill
}
\right.
\label{convective-discrete}
\end{equation}
where a linear approximation between discrete values $\Pi^{j-1}_i, i=0,1, ..., n,$ 
is being used to compute the value
$\Pi^{j-1}(x_i - \ln\ro_j + \ln\ro_{j-1} - (r-q)k)$.

The diffusive part can be solved numerically by means of finite differences. Using central finite difference
for approximation of the derivative $\partial_x\Pi^j$ we obtain

\begin{eqnarray}
&&\frac{\Pi_i^j - \Pi_i^{j-\onehalf}}{k} + r\Pi_i^j
- (\sigma^j_i)^2 \frac{\Pi_{i+1}^j-\Pi_{i-1}^j}{2h}
- \frac{1}{2 h} \left( 
(\sigma_i^j)^2 \frac{\Pi_{i+1}^j - \Pi_{i}^j}{h}
-
(\sigma_{i-1}^j)^2 \frac{\Pi_{i}^j - \Pi_{i-1}^j}{h}
\right)  =0\,.
\nonumber
\end{eqnarray}
Hence, the vector of discrete values $\Pi^j=\{\Pi_i^j, i=1,2, ..., n\}$ at the time level 
$j\in\{1,2, ..., m\}$ satisfies the tridiagonal system of equations
\begin{equation}
\alpha_i^j \Pi_{i-1}^j + \beta_i^j \Pi_{i}^j + \gamma_i^j \Pi_{i+1}^j = \Pi_i^{j-\onehalf}
\label{eq-tridiag}
\end{equation}
for $i=1,2, ..., n,$  where
\begin{eqnarray}
\label{abc}
\alpha_i^j &\equiv& \alpha_i^j(\ro^j, \Pi^j)= -\frac{k}{2h^2} (\sigma_{i-1}^j)^2
+ \frac{k}{2h} \frac{(\sigma_{i}^j)^2}{2}
\nonumber \\
\gamma_i^j &\equiv& \gamma_i^j(\ro^j, \Pi^j) = -\frac{k}{2h^2} (\sigma_{i}^j)^2
- \frac{k}{2h} \frac{(\sigma_i^j)^2}{2}
\\
\beta_i^j &\equiv& \beta_i^j(\ro^j, \Pi^j)=  1 + r k - (\alpha_i^j +\gamma_i^j)\,.
\nonumber
\end{eqnarray}
The initial and boundary conditions at $\tau=0$ and $x=0, L,$ resp., can be approximated as
follows:
\[
\Pi_i^0 = \left\{
\begin{array}{lll}
-E & \ \ \hbox{for} \ x_i <\ln\left({r/q}\right)\hfil
\\
\ \ \ 0 & \ \ \hbox{for} \ x_i \ge \ln\left({r/q}\right) \hfil
\end{array}
\right.
\]
for $i=0,1, ..., n,$ and $\Pi_0^j = -E, \quad \Pi_n^j = 0$.

Next we proceed by approximation of equation (\ref{ro}) which introduces a nonlinear
constraint condition between the early exercise boundary function $\ro(\tau)$ and
the trace of the solution $\Pi$ at the boundary $x=0$ ($S=S_f(t)$ in the 
original variable). Taking a finite difference approximation of $\partial_x\Pi$ at 
the origin $x=0$ we obtain
\begin{equation}
\ro^j = \frac{rE}{q} + \frac{1}{2q} \sigma^2 \left((\Pi_1^j-\Pi_0^j)/h, \ro^j, \tau\right)
\frac{\Pi_1^j-\Pi_0^j}{h}\,.
\label{eq-ro}
\end{equation}

Now, equations (\ref{convective-discrete}), (\ref{eq-tridiag}) and (\ref{eq-ro}) can be written 
in an abstract form as a system of nonlinear equations:
\begin{eqnarray}
\ro^j &&= {\mathcal F}(\Pi^j, \ro^j)\nonumber\\
\Pi^{j-\onehalf} &&={\mathcal T}(\Pi^j, \ro^j)\label{abstract}\\
{\mathcal A}(\Pi^j, \ro^j) \Pi^j &&= \Pi^{j-\onehalf}\nonumber
\end{eqnarray}
where 
${\mathcal F}(\Pi^j, \ro^j)$ is the right-hand side of the algebraic equation (\ref{eq-ro}),
${\mathcal T}(\Pi^j, \ro^j)$ is the transport equation solver given by 
the right-hand side of (\ref{convective-discrete}) and
${\mathcal A}={\mathcal A}(\Pi^j, \ro^j)$ is a tridiagonal matrix with coefficients given 
by (\ref{abc}). The system (\ref{abstract}) can be approximately solved by means of successive iterates 
procedure. We define, for $j\ge 1,$  $\Pi^{j,0} = \Pi^{j-1}, \ro^{j,0} = \ro^{j-1}$. Then the 
$(p+1)$-th approximation of $\Pi^j$ and $\ro^j$ is obtained as a solution to the system:
\begin{eqnarray}
\ro^{j,p+1} &&= {\mathcal F}(\Pi^{j,p}, \ro^{j,p})\nonumber\\
\Pi^{j-\onehalf, p+1} &&={\mathcal T}(\Pi^{j,p}, \ro^{j,p+1})\label{abstract-iter}\\
{\mathcal A}(\Pi^{j,p}, \ro^{j,p+1}) \Pi^{j,p+1} &&= \Pi^{j-\onehalf, p+1}\,.\nonumber
\end{eqnarray}
Notice that the last equation is a linear tridiagonal equation for the vector $\Pi^{j,p+1}$
whereas $\ro^{j,p+1}$ and $\Pi^{j-\onehalf, p+1}$ can be directly computed from (\ref{eq-ro}) and 
(\ref{convective-discrete}), resp.
Now, if the sequence of approximate solutions $\{(\Pi^{j,p}, \ro^{j,p}) \}_{p=1}^\infty$ converges to 
some limiting value $(\Pi^{j,\infty}, \ro^{j,\infty})$ then this limit is a solution to a nonlinear 
system of  equations (\ref{abstract}) at the time level $j$ and we can  proceed by computing 
the approximate solution the next time level $j+1$.

\section{Numerical approximations of the early exercise boundary}

In this section we focus on numerical experiments based on the iterative  
scheme described 
in the previous section. The main purpose is to compute the free boundary profile 
$S_f(t)=\ro(T-t)$ for different (non)linear Black--Scholes models and for various model
parameters. We used $n = 750$ spatial points and $m = 225000$ time discretization steps. 
In average we needed $p\le6$ micro-iterates (\ref{abstract-iter}) in order to solve the nonlinear 
system (\ref{abstract}) with the precision $10^{-7}$.
A solution $(\Pi,\ro)$ has been computed by our iterative algorithm for the following basic model parameters:
$E=10, T=1, r=0.1, q=0.05,$ and  $\hat\sigma=0.2$.

\subsection{Case of a constant volatility -- comparison study}

\begin{figure}

\begin{center}
\includegraphics[width=0.35\textwidth]{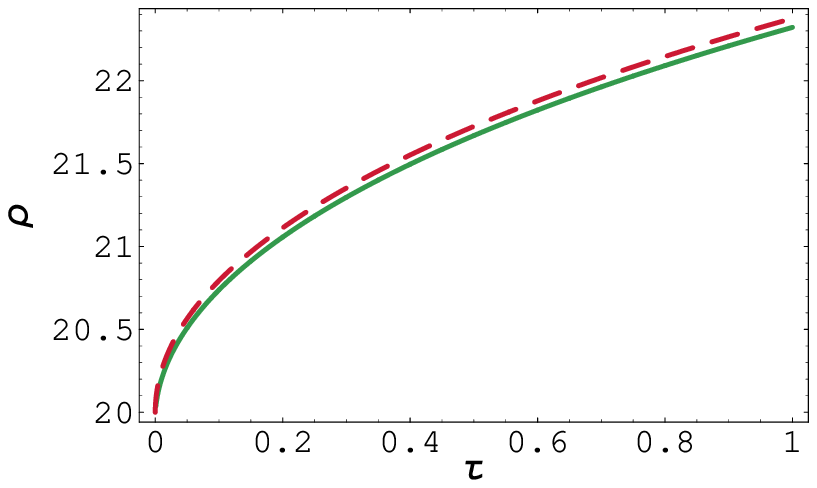}
\hglue 1truecm
\includegraphics[width=0.35\textwidth]{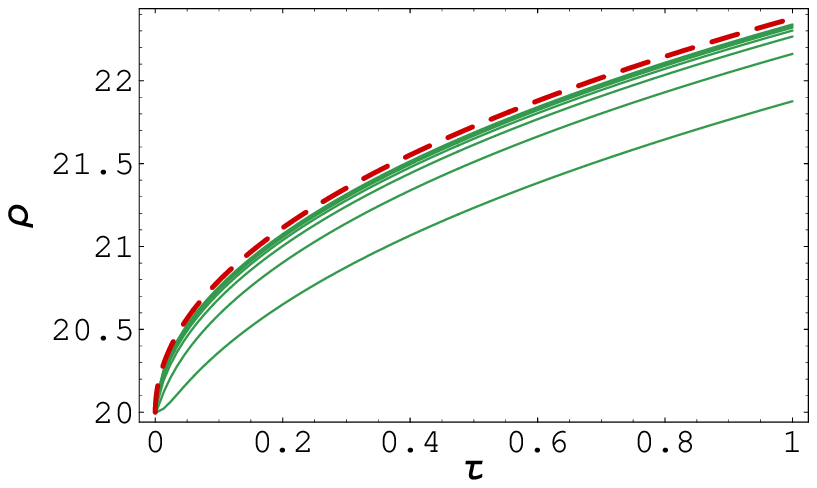}

\centerline{\small a) \hglue 0.4\textwidth b)}

\vglue 0.3truecm
\includegraphics[width=0.35\textwidth]{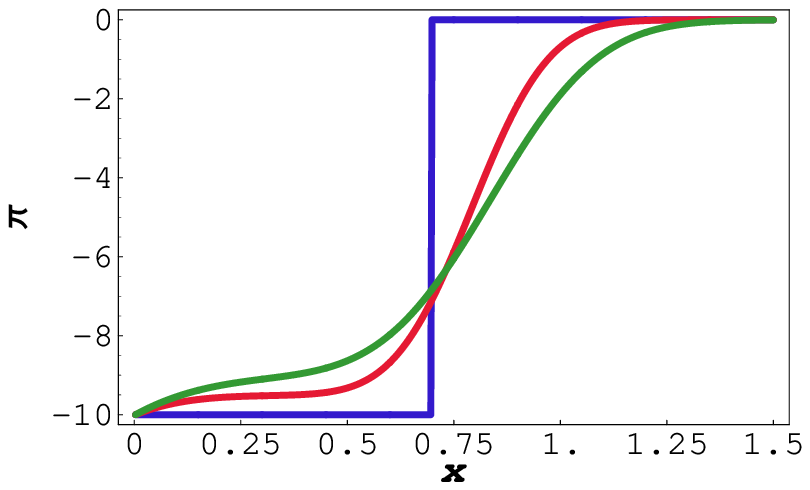}
\hglue 1truecm
\includegraphics[width=0.35\textwidth]{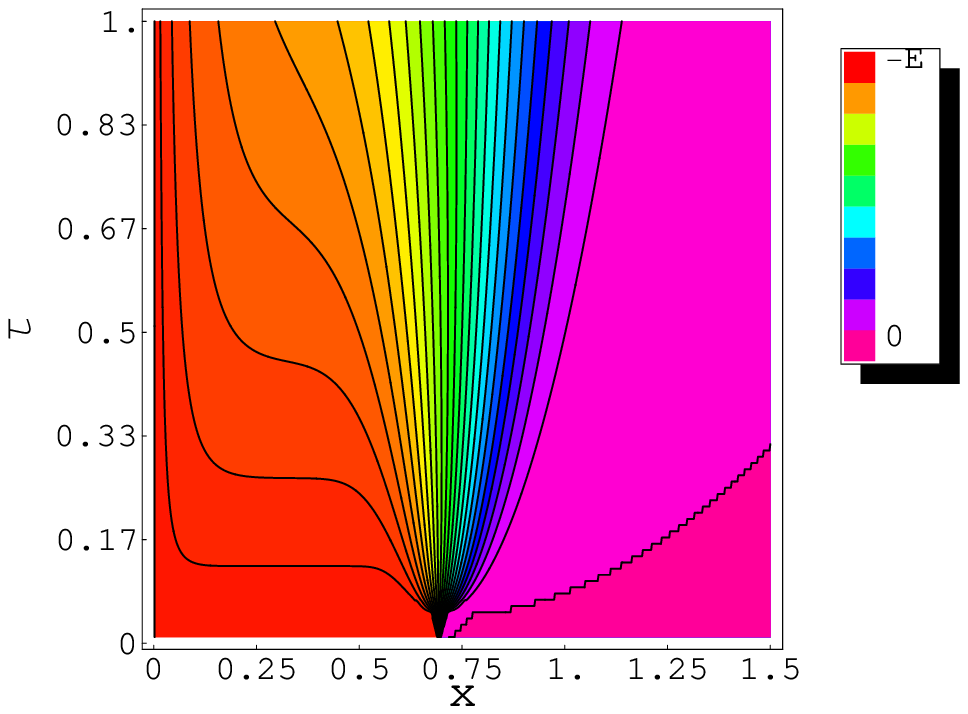}

\centerline{\small c) \hglue 0.4\textwidth d)}
\end{center}

\caption[]{\small
a) A comparison of the free boundary function $\ro(\tau)$ computed by the iterative algorithm 
(green solid curve) to the integral equation based  approximation (dashed red curve); 
b) free boundary positions computed for various mesh sizes;
c) a solution profile $\Pi(x,\tau)$ for $\tau=0$ (blue line), $\tau=T/2$ (red curve),
$\tau=T$ (green curve); d) contour plot of the function $\Pi(x,\tau)$.
}
\vspace{2mm}
\label{fig-noRAPM}
\end{figure}

In our first numerical experiment we make attempt to compare our iterative approximation scheme for 
solving the
free boundary problem for an American Call option to known schemes in the case when the volatility
$\sigma>0$ is constant. We compare our solution to the one computed by means of
a solution to a nonlinear integral equation for $\ro(\tau)$ as described in Remark~1 (see also 
\cite{Se,SSC}). This comparison can be also considered as a benchmark or test example
for which we know a solution that can be computed by a another justified algorithm. 
In Fig.~\ref{fig-noRAPM}, part a),  
we show the function $\ro$ computed by our iterative algorithm for  $E=10, T=1, r=0.1, 
q=0.05, \sigma=0.2$. 
At the expiry $T=1$ the value of $\ro(T)$ was computed as: $\ro(T)=22.321$. The corresponding value
$\ro(T)$ computed from the integral equation (see \cite{Se}) was  $\ro(T)=22.375$.
The relative error is less than 0.25\%. In the part b) we present 7 approximations of the free boundary function
$\ro(\tau)$ computed for different mesh sizes $h$ (see Tab.~\ref{tab1} for details). 
The sequence of approximate free boundaries $\ro_h, h=h_1, h_2, ... ,$ converges
monotonically from below to the free boundary function $\ro$ as $h\downarrow 0$.
The next part c) of Fig.~\ref{fig-noRAPM} depicts
various solution profiles of a function $\Pi(x,\tau)$. In order to achieve a good approximation
to equation (\ref{eq-ro}) we need very accurate approximation of $\Pi(x,\tau)$ for $x$ close to the
origin $0$. The last part d) of Fig.~\ref{fig-noRAPM} depicts the contour plot of the function 
$\Pi(x,\tau)$.

In Tab.~\ref{tab1} we present the numerical error analysis for the distance $\Vert\ro_h -\ro\Vert_p$
measured in two different norms ($L^\infty$ and $L^2$) 
of a computed free boundary position $\ro_h$ corresponding to the mesh size $h$ and the solution $\ro$
computed from the integral equation described in Remark~1 (see also \cite{Se}). The time step $k$ has been
adjusted to the spatial mesh size $h$ in order to satisfy CFL condition $\hat\sigma^2k/h^2 
\approx 1/2$.
We also computed the experimental order of convergence $\hbox{eoc}(L^p)$ for $p=2,\infty$. Recall that the experimental
order of convergence can be defined as the ratio
\[
\hbox{eoc}(L^p) = \frac{\ln(\Vert\ro_{h_i} -\ro\Vert_p) - \ln(\Vert\ro_{h_{i-1}} -\ro\Vert_p)}{\ln h_i - \ln h_{i-1}}\,.
\]
It can be interpreted as an exponent $\alpha=\hbox{eoc}(L^p)$ for which we have
$\Vert\ro_{h} -\ro\Vert_p = O(h^\alpha)$. It turns out from Tab.~\ref{tab1} that it is reasonable to
make a conjecture that $\Vert\ro_{h} -\ro\Vert_\infty = O(h)$ whereas $\Vert\ro_{h} -\ro\Vert_2 = O(h^{3/2})$
as $h\to 0^+$.

\begin{table}
\begin{center}
\small
\caption{\small Experimental order of convergence of the iterative algorithm for approximating the free
boundary position.}
\smallskip
\begin{tabular}
{l|l|l|l|l}
\hline
\hline
$h$& err($L^\infty$) & eoc($L^\infty$)& err($L^2$)& eoc($L^2$) \\
\hline
\hline
0.03& 0.5& -& 0.808& - \\
0.012& 0.215& 0.92& 0.227& 1.39 \\
0.006& 0.111& 0.96& 0.0836& 1.44 \\
0.004& 0.0747& 0.97& 0.0462& 1.46 \\
0.003& 0.0563& 0.98& 0.0303& 1.47 \\
0.0024& 0.0452& 0.98& 0.0218& 1.48 \\
0.002& 0.0378& 0.98& 0.0166& 1.48 \\
\hline
\end{tabular}
\label{tab1}
\end{center}
\end{table}

\subsection{Risk Adjusted Pricing Methodology model}

\begin{figure}

\begin{center}
\includegraphics[width=0.4\textwidth]{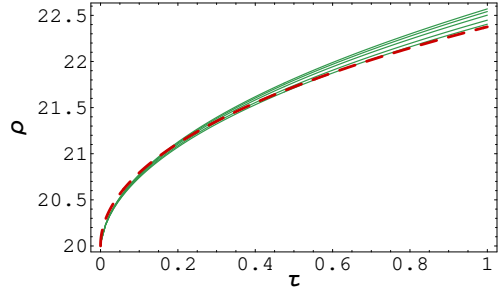}
\end{center}

\caption[]{\small
A comparison of the free boundary function $\ro^R(\tau)$ computed for the Risk Adjusted Pricing
Methodology model. Dashed red curve represents a solution corresponding to 
$R=0$, whereas the green curves represent a solution $\ro^R(\tau)$ 
for different values of the risk premium coefficients 
$R=5, 15, 40, 70, 100$.
}
\vspace{2mm}
\label{fig-rapm}
\end{figure}

In the next example we computed the position of the free boundary $\ro(\tau)$ in the case
of the Risk Adjusted Pricing Methodology model - a nonlinear Black--Scholes type model derived 
by Janda\v{c}ka and \v{S}ev\v{c}ovi\v{c} in \cite{JS}. In this model the volatility $\sigma$ is
a nonlinear function of the asset price $S$ and the second derivative $\partial_S^2 V$ of the 
option price, and it is given by  formula (\ref{RAPM}).
In Fig.~\ref{fig-rapm} we present results of numerical approximation of the free boundary position
$\ro^R(\tau)=S_f^R(T-\tau)$ in the case when 
the coefficient of transaction costs $C=0.01$ is fixed and the risk premium measure $R$ 
varies from 
$R=5, 15, 40, 70,$ up to $R=100$. We compare 
the position of the free boundary $\ro^R(\tau)$ to the case when there are no transaction costs
and no risk from volatile portfolio, i.e. we compare it with the free boundary position $\ro^0(\tau)$
for the linear Black--Scholes equation (see Fig.~\ref{fig-rapm}). 
An increase in the  risk premium coefficient $R$ resulted in an increase of the free boundary position 
as it can be expected.

\begin{table}
\begin{center}
\small
\caption{\small 
Distance  $\Vert \ro^R-\ro^0\Vert_p$ ($p=2,\infty$) of the free boundary position 
$\ro^R$ from the reference free boundary position $\ro^0$ and orders $\alpha_\infty$
and $\alpha_2$ of approximation.}
\smallskip
\begin{tabular}
{r|l|l|l|l}
\hline
\hline
$R$& $\Vert \ro^R-\ro^0\Vert_\infty$ & $\alpha_\infty$ & $\Vert \ro^R-\ro^0\Vert_2$& $\alpha_2$ \\
\hline
\hline
1 & 0.0601& -& 0.0241& -\\
2 & 0.0754& 0.33& 0.0303& 0.328\\
5 & 0.102& 0.33& 0.0408& 0.326\\
10 & 0.128& 0.33& 0.0511& 0.324\\
15 & 0.145& 0.32& 0.0582& 0.323\\
20 & 0.16& 0.32& 0.0639& 0.322\\
30 & 0.182& 0.32& 0.0727& 0.321\\
40 & 0.2& 0.32& 0.0798& 0.32\\
50 & 0.214& 0.32& 0.0856& 0.319\\
60 & 0.227& 0.32& 0.0907& 0.318\\
70 & 0.239& 0.32& 0.0953& 0.317\\
80 & 0.249& 0.32& 0.0994& 0.317\\
90 & 0.259& 0.32& 0.103& 0.316\\
100 & 0.268& 0.32& 0.107& 0.316\\
\hline
\end{tabular}
\label{tab2}
\end{center}
\end{table}

\begin{figure}

\begin{center}
\includegraphics[width=0.35\textwidth]{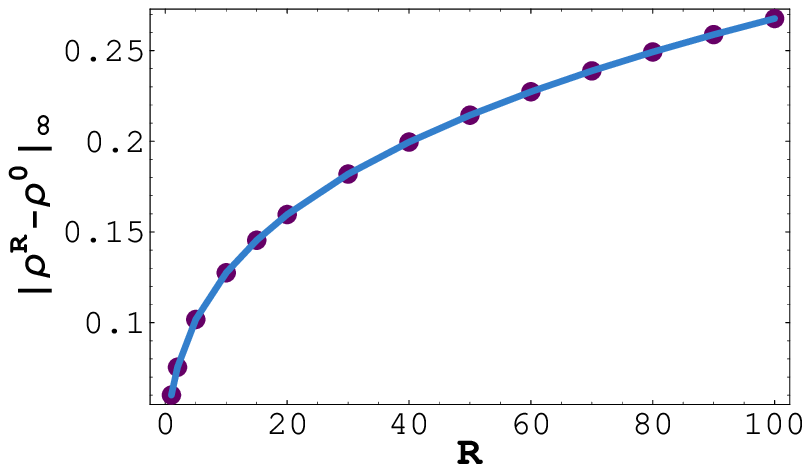}
\includegraphics[width=0.35\textwidth]{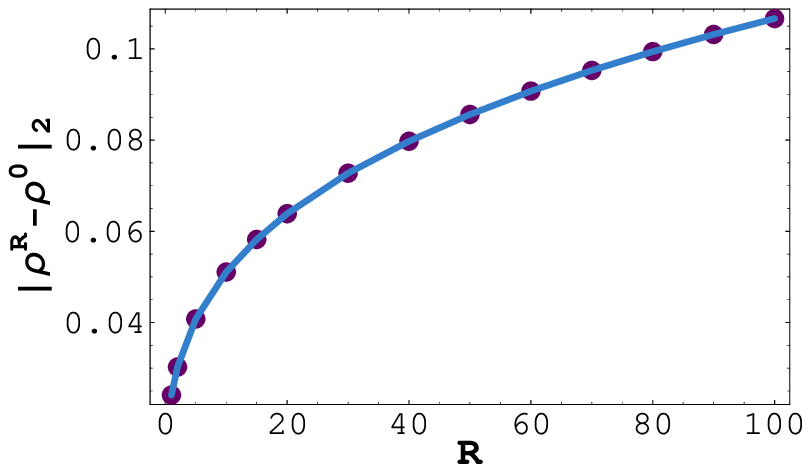}
\end{center}

\caption[]{\small
Dependence of the norms $\Vert \ro^R -\ro^0\Vert_p$ ($p=\infty, 2$) of the deviation of the free boundary 
$\ro=\ro^R(\tau)$ for the RAPM model on the risk premium coefficient $R$. 
}
\vspace{2mm}
\label{fig-rapm-err}
\end{figure}

In Tab.~\ref{tab2} and Fig.~\ref{fig-rapm} 
we summarize results of comparison of the free boundary position $\ro^R$ for various values
of the risk premium coefficient to the reference
position $\ro=\ro^0$ computed from the Black--Scholes model with constant volatility $\sigma=\hat\sigma$, i.e.
$R=0$. The experimental order $\alpha_p$ of the distance function 
$\Vert \ro^R-\ro^0\Vert_p = O(R^{\alpha_p})$ has been computed for $p=2,\infty$, 
as follows:
\[
\alpha_p = \frac{\ln(\Vert\ro^{R_i} -\ro^0\Vert_p) - \ln(\Vert\ro^{R_{i-1}} -\ro^0\Vert_p)}{\ln R_i - \ln R_{i-1}}\,.
\]
According to the values presented in Tab.~\ref{tab2} it turns out that
the plausible conjecture is that $\Vert \ro^R-\ro^0\Vert_p = O(R^{1/3})$ 
for both norms $p=2$ and $p=\infty$. Since the transaction cost coefficient $C$ and risk premium
measure $R$ enter the
expression for the RAPM volatility (\ref{RAPM}) only in the product $C^2 R$ we can conjecture
that $\Vert \ro^{R,C}-\ro^{0,0}\Vert_p = O(C^{2/3} R^{1/3})$ as either $C\to 0^+$ or $R\to 0^+$.

\subsection{Barles and Soner model}

\begin{figure}

\begin{center}
\includegraphics[width=0.4\textwidth]{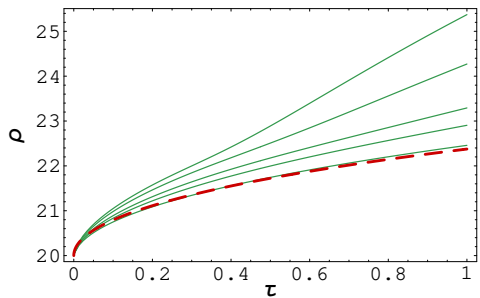}
\end{center}

\caption[]{\small
A comparison of the free boundary function $\ro(\tau)$ computed for the Barles and Soner
model. Dashed red curve represents a solution corresponding to 
$R=0$, whereas the green curves represents a solution $\ro(\tau)$ 
for different values of the risk aversion coefficient 
$a=0.01, 0.07, 0.13, 0.25, 0.35$.
}
\vspace{2mm}
\label{fig-barles}
\end{figure}

\begin{table}
\begin{center}
\small
\caption{\small 
Distance  $\Vert \ro^a-\ro^0\Vert_p$ ($p=2,\infty$) of the free boundary position 
$\ro^a$ from the reference free boundary position $\ro^0$ orders $\alpha_\infty$
and $\alpha_2$ of approximation.}
\smallskip
\begin{tabular}
{l|l|l|l|l}
\hline
\hline
$a$& $\Vert \ro^a-\ro^0\Vert_\infty$ & $\alpha_\infty$ & $\Vert \ro^a-\ro^0\Vert_2$& $\alpha_2$ \\
\hline
\hline
0.01& 0.156& -& 0.0615& -\\
0.02& 0.25& 0.68& 0.0985& 0.68\\
0.05& 0.472& 0.69& 0.184& 0.679\\
0.07& 0.602& 0.72& 0.232& 0.69\\
0.1& 0.793& 0.77& 0.298& 0.712\\
0.11& 0.857& 0.82& 0.32& 0.74\\
0.13& 0.99& 0.86& 0.364& 0.766\\
0.15& 1.13& 0.92& 0.409& 0.807\\
0.2& 1.52& 1.& 0.529& 0.897\\
0.25& 1.97& 1.2& 0.669& 1.05\\
0.3& 2.49& 1.3& 0.833& 1.21\\
0.35& 3.07& 1.4& 1.03& 1.35\\
\hline
\end{tabular}
\label{tab3}
\end{center}
\end{table}

The last example is devoted to the nonlinear Black--Scholes model due to Barles and Soner 
(see \cite{BaSo}). In this model the volatility is given by equation (\ref{barles}). 
Numerical results are depicted in 
Fig.~\ref{fig-barles}. Choosing a larger value of the risk aversion 
coefficient $a$ resulted in increase of the free boundary position $\ro^a(\tau)$.
The position of the early exercise boundary $\ro^a(\tau)$ has 
considerably increased in comparison to the linear Black--Scholes equation with constant
volatility $\sigma=\hat\sigma$. In contrast to the case of constant volatility as well as
the RAPM model, there is, at least a numerical evidence (see Fig.\ref{fig-barles} and $\ro^a$
for the largest $a=0.35$) that 
the free boundary profile $\ro^a(\tau)$ need not be necessarily convex.
Recall that that convexity of the free boundary profile has been proved analytically
by Ekstr\"om  {\em et al.} and Chen  {\em et al.} in a recent papers \cite{CC,E,ET} in the case of a American Put option and constant 
volatility $\sigma=\hat\sigma$.

\begin{figure}

\begin{center}
\includegraphics[width=0.35\textwidth]{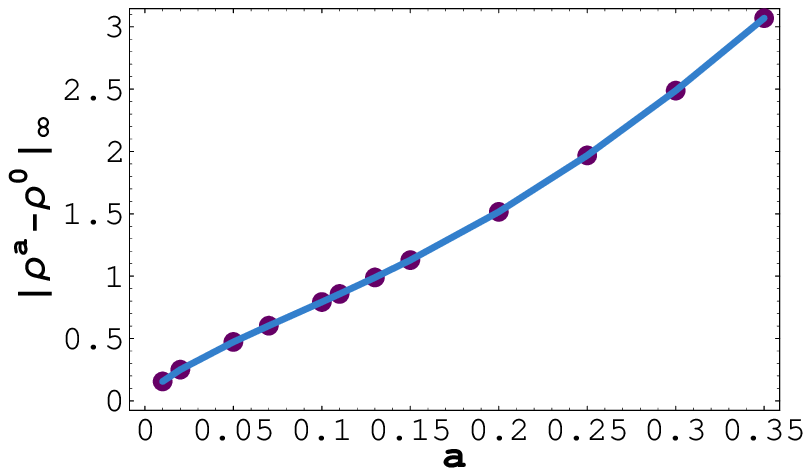}
\includegraphics[width=0.35\textwidth]{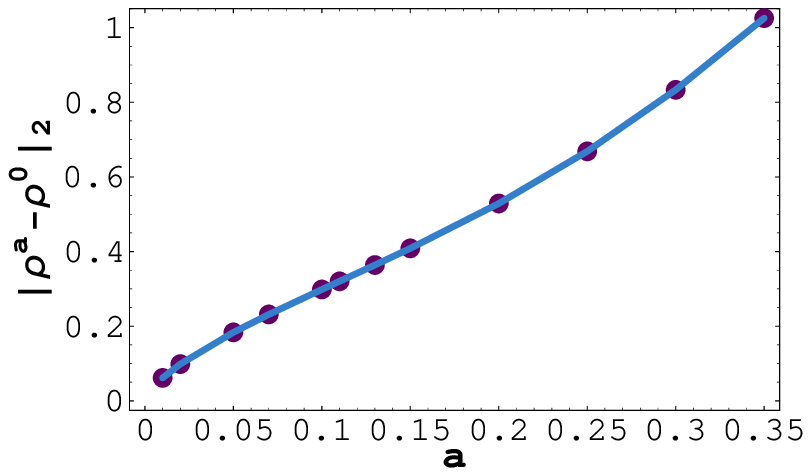}
\end{center}

\caption[]{\small
Dependence of the norms $\Vert \ro^a -\ro^0\Vert_p$ ($p=\infty, 2$) of the deviation of the free boundary 
$\ro=\ro^a(\tau)$ for the Barles-Soner model on the risk aversion parameter $a$. 
}
\vspace{2mm}
\label{fig-barles-err}
\end{figure}

Similarly as in the previous model we also investigated  the dependence of the free boundary 
position $\ro=\ro^a(\tau)$ on the risk aversion parameter $a>0$. 
In Tab.~\ref{tab3} and Fig.~\ref{fig-barles-err} we present results of comparison of the free boundary position $\ro^a$ for various values
of the risk aversion coefficient $a$ to the reference
position $\ro=\ro^0$. Inspecting values $\alpha_p$ of the order of distance $\Vert \ro^a-\ro^0\Vert_p$ 
it can be conjectured that $\Vert \ro^a-\ro^0\Vert_p = O(a^{2/3})$ as $a\to 0$.
for both norms $p=2$ and $p=\infty$.

\section{Conclusions}

We proposed a new iterative numerical scheme for approximating of the
early exercise boundary for a class of Black--Scholes equations for 
pricing American options with a volatility nonlinearly depending in the 
asset prices and the second derivative of the option price. The method consisted
of transformation the free boundary problem for the early exercise boundary 
position into a solution of a nonlinear parabolic equation and a nonlinear
algebraic constraint equation. The transformed problem has been solved by means
of operator splitting iterative technique. We also presented results of
numerical approximation of the free boundary for several nonlinear
Black--Scholes equation including, in particular, Barles and Soner model and
the Risk adjusted pricing methodology model.


\begin{thebibliography}{9}

\bibitem {A}
\textsc{Alobaidi, G.} (2004)
Laplace transforms and installment options.
\textit{Math. Models \& Methods in Appl. Science},
\textbf{18} (8), 1167--1189.

\bibitem{AP} 
\textsc{Avellaneda, M. \& Paras, A.} (1994),
Dynamic Hedging Portfolios for Derivative Securities in the Presence of
Large Transaction Costs.
\textit{Applied Mathematical Finance},
\textbf{1,} 165--193.


\bibitem{BW}
\textsc{Barone-Adesi, B. \&  Whaley, R. E.} (1987)
Efficient analytic approximations of American option values.
\textit{J. Finance},
\textbf{42}, 301--320.

\bibitem {BS}
\textsc{Black, F. \& Scholes, M.}  (1973)
The pricing of options and corporate liabilities.
\textit{J. Political Economy},
\textbf{81}, 637--654.


\bibitem {BaSo}
\textsc{Barles, G. \& Soner, H. M.} (1998)
Option Pricing with transaction costs and a nonlinear Black--Scholes equation.
\textit{Finance Stochast.},
\textbf{2}, 369-397.


\bibitem {CC}
\textsc{Xinfu Chen, Chadam, J., Lishang Jiang, \& Weian Zheng} (2006)
Convexity of the Exercise Boundary of the American Put Option on a Zero Dividend Asset,
\textit{Mathematical Finance}, to appear.


\bibitem{DH}
\textsc{Dewynne, J. N., Howison, S. D., Rupf, J. \& Wilmott, P.} (1993)
Some mathematical results in the pricing of American options.
\textit{Euro. J. Appl. Math.},
\textbf{4}, 381--398.

\bibitem {DFJ}
\textsc{During, B., Fournier, M. \& Jungel, A.} (2003)
High order compact finite difference schemes for a nonlinear Black--Scholes equation.
\textit{Int. J. Appl. Theor. Finance},
\textbf{7}, 767--789.

\bibitem {EM}
\textsc{Ehrhardt, M. \& Mickens, R.} (2006)
Discrete artificial boundary conditions for the Black--Scholes equation of American options, submmitted.

\bibitem {E}
\textsc{Ekstr\"om, E. \& Tysk, J.} (2006)
The American put is log-concave in the log-price.
\textit{Journal of Mathematical Analysis and Appl.},
\textbf{314}(2), 710--723.

\bibitem {ET}
\textsc{Ekstr\"om, E.} (2004)
Convexity of the optimal stopping boundary for the American put option
\textit{Journal of Mathematical Analysis and Appl.},
\textbf{299}(1), 147--156.

\bibitem {FP}
\textsc{Frey, R. \& Patie, P.} (2002)
Risk Management for Derivatives in Illiquid Markets:
A Simulation Study.
\textit{Advances in Finance and Stochastics},
Springer, Berlin, 137--159.

\bibitem {FS}
\textsc{Frey, R. \& Stremme, A.} (1997)
Market Volatility and Feedback Effects from Dynamic Hedging.
\textit{Mathematical Finance},
\textbf{4}, 351--374.

\bibitem {GJ}
\textsc{Geske, R. \& Johnson, H. E.} (1984)
The American put option valued analytically.
\textit{J. Finance},
\textbf{39}, 1511--1524.

\bibitem {GR}
\textsc{Geske, R.  \& Roll, R.} (1984)
On valuing American call options with the Black--Scholes
European formula.
\textit{J. Finance},
\textbf{89}, 443--455.

\bibitem {H}
\textsc{Hull, J.} (1989)
\textit{Options, Futures and Other Derivative Securities}, Prentice Hall.

\bibitem{HWW} 
\textsc{Hoggard, T., Whalley, A.E. \& Wilmott, P.} (1994)
Hedging option portfolios in the presence of transaction costs.
\textit{Advances in Futures and Options Research},
\textbf{7}, 21--35.

\bibitem{JS}
\textsc{Janda\v{c}ka, M., \& \v{S}ev\v{c}ovi\v{c}, D.} (2005)
On the risk adjusted pricing methodology based valuation of vanilla options
and explanation of the volatility smile.
\textit{Journal of Applied Mathematics},
\textbf{3}, 235--258.

\bibitem {KK}
\textsc{Kuske R. A. \& Keller, J. B.} (1998)
Optimal exercise boundary for an American put option.
\textit{Applied Mathematical Finance},
\textbf{5}, 107--116.

\bibitem{Kr} 
\textsc{Kratka, M.} (1998)
No Mystery Behind the Smile,
\textit{Risk},
\textbf{9,} 67--71.

\bibitem {Kw}
\textsc{Kwok, Y. K.} (1998)
\textit{ Mathematical Models of Financial Derivatives}.
Springer-Verlag.

\bibitem{Le} 
\textsc{Leland, H. E.} (1985)
Option pricing and replication with transaction costs
\textit{Journal of Finance},
\textbf{40}, 1283--1301.

\bibitem {Mac}
\textsc{MacMillan, L. W.} (1986)
Analytic approximation for the American put option.
\textit{ Adv. in Futures Options Res.},
\textbf{1}, 119--134.

\bibitem {Ma}
\textsc{Mallier, R.} (2002)
Evaluating approximations for the American put option. 
\textit{Journal of Applied Mathematics},
\textbf{2}, 71--92.

\bibitem {MA}
\textsc{Mallier, R. \& Alobaidi, G.} (2004)
The American put option close to expiry.
\textit{Acta Mathematica Univ. Comenianae},
\textbf{73}, 161--174.

\bibitem {M}
\textsc{Mynemi, R.} (1992)
The pricing of the American option.
\textit{ Annal. Appl. Probab.},
\textbf{2}, 1--23.


\bibitem {R}
\textsc{Roll, R.} (1977)
An analytic valuation formula for unprotected American call options on
stock with known dividends.
\textit{J. Finan. Economy},
\textbf{5}, 251--258.


\bibitem {Se}
\textsc{\v{S}ev\v{c}ovi\v{c}, D.} (2001)
Analysis of the free boundary for the pricing of an American call option.
\textit{Euro. Journal on Applied Mathematics},
\textbf{12}, 25--37.

\bibitem {SW}
\textsc{Sch\"onbucher, P., \&  Wilmott, P.} (2000)
The feedback-effect of hedging in illiquid markets.
\textit{SIAM Journal of Applied Mathematics}, 
\textbf{61}, 232--272.



\bibitem {SSC}
\textsc{Stamicar, R., \v{S}ev\v{c}ovi\v{c}, D. \& Chadam, J.} (1999)
The early exercise boundary for the American put near expiry:
numerical approximation.
\textit{Canad. Appl. Math. Quarterly},
\textbf{7}, 427--444.

\bibitem {ZCD}
\textsc{Jichao Zhao, Corless, R. M., \& Davison, M.} (2005)
Compact finite difference method for American option pricing.
\textit{Journal of Computational and Applied Mathematics}, to appear, 
available online 22 August 2006. 



\end{thebibliography}
\end{document}